\documentstyle[11pt,blois2002,epsfig]{article}

\def\Journal#1#2#3#4{{#1} {\bf #2}, #3 (#4)}

\def\NPB{{\em Nucl. Phys.} B}
\def\PLB{{\em Phys. Lett.}  B}

\def\PRD{{\em Phys. Rev.} D}

 \def\beq{\begin{equation}}
\def\eeq{\end{equation}}
\def\berr{\begin{eqnarray}}
\def\eerr{\end{eqnarray}}

\begin{document}

\title{Antimatter regions in the baryon-dominated Universe}

\author{Maxim~Yu.~Khlopov$^{*\ \#}$, Sergei~G.~Rubin$^{*\ \#}$ and Alexander~S.~Sakharov$^{\dag\ \ddag}$}

\address{$^{*}$ Moscow Engineering Physics Institute, Kashirskoe shosse 31, 115409
Moscow,  Russia \\
$^{\#}$ Center for CosmoParticle Physics "Cosmion", Miussk. pl. 4, 125047
Moscow, Russia\\$^{\dag}$ Theory Division, CERN, CH-1211 Geneva 23, Switzerland\\
$^{\ddag}$ Swiss Institute of Technology, ETH-Z\"urich, 8093 Z\"urich, Switzerland\\
{\rm XIVth RENCONTRES DE BLOIS, MATTER--ANTIMATTER ASYMMETRY}
}

\maketitle

\begin{abstract}

Quantum fluctuations
of a complex, baryonic charged scalar field caused by inflation can
generate  large domains, which convert later into antimatter regions. As a
result the Universe can become globally matter-dominated, with minor contribution
of antimatter regions. The distribution and evolution of such antimatter regions
could cause every galaxy to be a harbour of an  anti-star globular cluster. At
the same time, the scenario does not lead to large-scale isocuvature
perturbations, which would disturb observable CMB anisotropy. The existence of
one of such antistar globular cluster in our Galaxy does not contradict the
observed $\gamma$-ray background, but the expected fluxes of
$\overline{\rm ^4He}$ and $\overline{\rm ^3He}$ from such an antimatter object are
definitely accessible to the sensitivity of the coming AMS--02 experiment.

\end{abstract}

\section{Introduction}
The whole set of astrophysical observations \cite{exl} prefers that our Universe
be globally matter/antimatter asymmetrical. This statement comes mostly from
the fact that  equal amounts of matter and antimatter domains, coexisting with
each other, would  annihilate on their borders, consequently disturbing the
observed diffused $\gamma$-ray background \cite{exl,crg}. A closed contact of
coexisting matter and antimatter at the early epochs is almost unavoidable
\cite{kolb}. The $\gamma$-ray flux at $100$ MeV range caused by this kind of
annihilation would be below the observable one only in the case when the
characteristic size of domains exceeds $10^3$ Mpc \cite{crg}. This fact requires baryon domination over the whole volume
of the Universe.

  However, the above mentioned arguments cannot exclude the Universe composed
almost entirely of matter, with relatively small insertions of antimatter
regions. The fate of such antimatter regions depends on their size. If the
physical size of some of them is larger than  the critical surviving size
$L_c=8h^2$ kpc~\cite{we}, they survive annihilation with surrounding matter. It is very likely that the
dense fraction of the antimatter domains out of the preserved population evolve into condensed antimatter
astrophysical objects \cite{khl}.

In this report we consider the model of inhomogeneous baryogenesis \cite{zil}
based on the inflationary evolution of the baryon charged scalar field. This
scenario makes it reasonable to discuss the existence of an antistar globular
cluster (GC) in our Galaxy, preventing at the same time large-scale isocurvature fluctuations,
which could be imprinted into CMB anisotropy.
The main experimental signature of the discussed scenario is indicated in
the expected fluxes of  $\overline{\rm ^4He}$ and $\overline{\rm ^3He}$,
which are accessible for the sensitivity of AMS--02 detector \cite{ams}.

\section{Scenario of inhomogeneous baryogenesis with antimatter generation}
\begin{figure} [t] \begin{center}
\epsfig{file=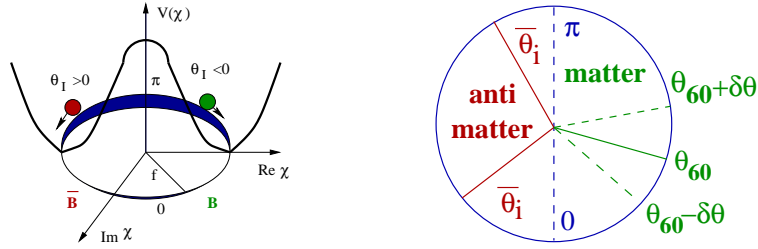,width=100mm,clip=} \end{center}
\vspace*{-0.1cm}\caption{{\bf Left panel:} PNG potential in the spontaneous
baryogenesis mechanism. The sign of produced baryon asymmetry depends on the
starting point of oscillations. {\bf Right panel:} The inflational evolution of
the phase. The phase  $\theta_{60}$ sits in the range $[\pi ,0 ]$ at the
beginning of inflation and makes Brownian step
$\delta\theta_{eff}=H_{infl}/(2\pi f_{eff})$  at each e--fold. The typical
wavelength of the fluctuation $\delta\theta$ is equal to $H^{-1}_{infl}$. The
whole domain  $H^{-1}_{infl}$, containing phase $\theta_{N}$ gets  divided,
after one e--fold, into  $e^3$ causally disconnected domains of radius
$H^{-1}_{infl}$. Each new domain contains almost homogeneous phase value
$\theta_{N-1}=\theta_{N}\pm\delta\theta_{eff}$. Every successive e-fold this
process repeats in every domain.} \label{spont}\vspace*{-0.5cm} \end{figure}

 Our approach \cite{zil} is  based on the
spontaneous baryogenesis mechanism \cite{sb}, which implies the existence
of a complex scalar field $\chi =(f/\sqrt{2})\exp{(\theta )}$ carrying
the baryonic charge. The $U(1)$ symmetry, which corresponds to the baryon
charge, is broken spontaneously and explicitly. The explicit breakdown of $U(1)$
symmetry is caused by the phase-dependent term
 \beq\label{expl} V(\theta )=\Lambda^4(1-\cos\theta ),
 \eeq
which results in the pseudo Nambu--Goldstone (PNG) potential of
Fig.~\ref{spont}. The possible lepton-number violating interaction of the field
$\chi$ with matter fields can have the following structure \cite{dolgmain,zil}
\beq\label{leptnumb} {\cal L}=g\chi\bar QL+{\rm h.c.}, \eeq
where fields $Q$ and $L$ represent a heavy quark and lepton, coupled to the
ordinary matter fields. In the early Universe, at a time when the friction term,
induced by the Hubble constant, becomes comparable with the angular mass
$m_{\theta}=\frac{\Lambda^2}{f}$, the phase $\theta$ starts to oscillate around
the minima of the PNG potential and decays into matter fields according to
(\ref{leptnumb}). The coupling (\ref{leptnumb}) gives rise to the following
\cite{dolgmain,zil}: as the phase starts to roll down in the clockwise direction
during the first oscillation (Fig.~\ref{spont}),  it preferentially creates
baryons over antibaryons, while the opposite is true as it starts to roll down
in the opposite direction. The baryon/antibaryon number, created in these
oscillations, is given by \cite{zil} \beq \label{baryonnumber} N_{B(\bar
B)}\approx\frac{g^2f^2m_{\theta}}{8\pi^2}{\cal
W}_{\theta_i}\theta_i^2\int\limits_{\mp\theta_i/2}^{\infty}d\omega\frac{\sin^2\omega}{\omega^2}, \eeq where ${\cal W}_{\theta_i}$ is the volume, in which the
phase has the value $\theta_i$. Thus, the distribution of the resulting baryon
charge reflects the primordial distribution of the phase $\theta$ in the early
Universe.

We suppose \cite{zil} that the radial mass $m_{\chi}$ of the field $\chi$ is
larger than the Hubble constant $H_{infl}$ during inflation, while for the
angular mass of $\chi$ just the opposite condition, $m_{\theta}\ll H_{infl}$, is
satisfied in that period. Thus $U(1)$ symmetry is already broken spontaneously
on the energy scale $f$ at the beginning of inflation, whereas the phase $\theta$ behaves like a massless scalar field.
This means that the quantum fluctuations of $\theta$ at the de
Sitter background \cite{linde} will define the primordial phase distribution in the early Universe.
Thus to have
a globally baryon-dominated Universe one must have  the phase sited in the
range $[\pi ,0 ]$ (Fig.~\ref{spont}), just at the beginning of inflation
\footnote{We put the duration of the inflation period to 60 e-folds.}  (when the
size of the modern Universe crosses the horizon). Then quantum fluctuations  in
some regions move the phase to the values $\bar\theta_i$ in the range $[0,\pi ]$
(Fig.~\ref{spont}, right panel) where a successive antibaryon excess gets
produced. If a domain with  $\bar\theta_i$ leaves the horizon
$H_{infl}^{-1}$ before the 45th e-fold \cite{zil}, it becomes biger than the
critical survival size $L_c$ and survives annihilation.

Since baryon/antibaryon numbers are produced by the out-of-equilibrium decay of
non-inflaton field $\chi$, the isocurvature perturbations \cite{riotto} will be
imprinted in baryons, giving rise to the effect on the CMB angular power
spectrum. The size of the effect is defined dispersion  by the $\delta\theta
=H_{infl}/(2\pi f)$ and diminished by a factor $\Omega_B /\Omega_{tot}$. At the
large scales corresponding to the first 6 e-folds of inflation, the measured CMB
anisotropy does not allow to be the dispersion larger, than $\delta\theta\le
10^{-3}$, while at the  $N_c$ e-fold corresponding to $L_c$ one needs to get the
phase already to the antibaryon production region of vacuum manifold, which
requires quite a large magnitude of dispersion  $\delta\theta\simeq 10^{-2}$
\cite{zil}. An outcome of this disagreement could be found by making the
dispersion $\delta\theta$ dynamically changing with respective to the current
e-fold. One assumes a vacuum that dynamics makes the energy scale of
$U(1)$ symmetry breaking $f$ variable during inflation. Phenomenologicaly the
requirement dynamics can be obeyed by the potential with the following couplings
\footnote{The coupling in (\ref{coupling}) can be generated by SUSY/SUGRA
potentials (see for details \cite{riotto})} of inflaton $\phi$ to $\chi$ \beq
\label{coupling} V(\phi ,\chi
)=\frac{1}{2}m_{\phi}\phi^2+\lambda\left(\chi\chi^*-\frac{f^2}{2}\right)^2-g_{\phi\chi}\chi\chi^*(\phi -cM_{Pl})^2. \eeq
The radius of phase vacuum manifold becomes e-fold dependent, making the
effective dispersion $\delta\theta_{eff}$ small at the beginning of inflation,
thereby suppressing the magnitude of large-scale isocurvature fluctuations
\beq
\label{radiuseffective}
f_{eff}(N)=f\sqrt{1+\frac{g_{\phi\chi}M_{Pl}}{12\pi\lambda}(N_c-N)};\
\delta\theta_{eff}=\frac{H_{inf}}{2\pi f_{eff}}. \eeq
The dispersion grows up to its maximun value, when inflantion reachs the $N_c$th e-fold. Then it decreases
again to a negligible value. Such dynamics allows to generate an above-critical
size progenitor of antimatter region in every volume box corresponding to each
galaxy, preserving at the same time the general barion asymmetry of the Universe
as a whole (see Fig.~\ref{antiheliumsignal}).  
Fig.\ref{antiheliumsignal}.
 The evolution of created baryon number density is straightforward \cite{zil}.
The other advantage of mechanism (\ref{coupling}), (\ref{radiuseffective}) is to
make the antimatter domain size distribution almost independent of the initial
position of the phase at the beginning of inflation, requiring only that it be
located somewhere in the baryon-production region of the vacuum manifold.

\section{Evolution and observational signature of antimatter domains}
The antibaryon number (\ref{baryonnumber}) in progenitors shows a strongly
rising dependence on the initial value of the phase $\bar\theta_i$, which  makes
sense to discuss the possibility of having high density antimatter region in
every galaxy. Let us consider the evolution of such a high density antimatter
region in the  surrounding matter.

It is well known \cite{glob} that a cloud of mass
$10^5M_{\odot}$--$10^6M_{\odot}$, which has temperature near $10^4$~K and a
density several tens of times that of the  surrounding hot gas, is
gravitationally unstable. This object is identified as a proto-object of GC and
reflects the Jeans mass at the recombination epoch. Thus if the phase
$\bar\theta_i$ inside an antimatter region progenitor was in a position to
generate an  antimatter density higher than the
surrounding matter density by one order of magnitude, it is very likely that
the region evolves into an antimatter GC \cite{khl}. GCs are the
oldest galactic star systems to form in the Universe, and contain stars of the
first population. Thereby a GC at large galactocentric distance is the ideal
astrophysical object that could be made out of antimatter.  The $\bar p$
releasing from such an antistar GC by the stellar wind and anti-supernova
explosions will be collected in our Galaxy and annihilate with $p$
giving a contribution into GeV range diffused $\gamma$-ray
background \cite{khlopgolubkov}. This contribution, being compared with the $\gamma$-- ray background
measured by EGRET  sets the upper limit on the mass of
antistar GC in our galaxy to $10^5M_{\odot}$ \cite{khlopgolubkov}, while $L_c$
defines the lower mass limit $10^3M_{\odot}$ on a possible antistar GC.
\begin{figure} [t] \begin{center} \epsfig{file=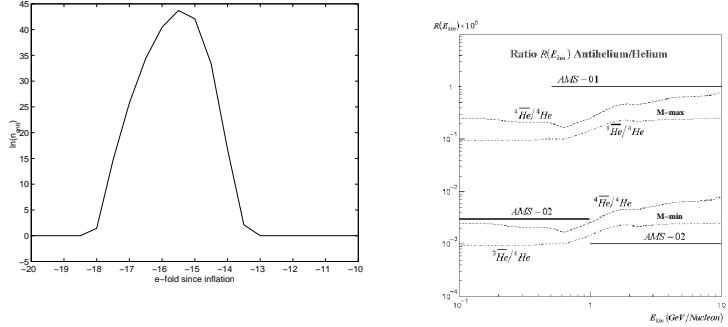,width=100mm,clip=}
\vspace*{-0.1cm}\caption{{\bf Left panel}: The size (e-fold) distribution of
antimatter domain progenitors calculated with respective to the variable
dispersion mechanism.  The calculations have been done under the assumptions
$H_{infl}=10^{13}$GeV, $\frac{g_{\phi\chi}M_{Pl}}{12\pi\lambda}\simeq 10^3$.
About $ 10^{11}$ of critical size antimatter domains appear, while the number of
larger domain as well as much smaller domains is highly suppressed. The volume occupated by antimatter is
less then $10^{-4}$ of the total volume of the Universe. {\bf Right panel}: The expected fluxes
of $\overline{^4He}$ and $\overline{^3He}$ from anti--star GC in the mass range
$10^3M_{\odot}$--$10^5M_{\odot}$, (M-min--M-max), in the comparison with the AMS02 sensitivity.}
\label{antiheliumsignal} \end{center}\vspace*{-0.5cm} \end{figure}

The most important experimental signature of the existence of an antistar GC in
our Galaxy, would be the observation of antinuclei in the cosmic rays near
the Earth's orbit \cite{bgk}. The expected fluxes of $\overline{\rm ^4He}$ and
$\overline{\rm ^3He}$ (Fig.~\ref{antiheliumsignal}) from such an antimatter
object \cite{bgk} are only a factor of 2 below the limit of the AMS--01
(STS--91) experiment \cite{amsl} and definitely accessible for the sensitivity
of the coming AMS--02 experiment~\cite{ams}.

\noindent {\bf \underline{Acknowledgement}} The speaker, A.~Sakharov, is
grateful to the Organizing Committe of the conference for their invitation and
financial support.

\end{document}